\documentclass[twocolumn,prl]{revtex4}
\usepackage{graphicx,amsmath}

\begin{document}

\title{Continuous-Variable Quantum State Transfer with Partially Disembodied Transport }

\author{Jing Zhang$^{\dagger} $, Changde Xie, Kunchi Peng}

\affiliation{State Key Laboratory of Quantum Optics and\\
Quantum Optics Devices, Institute of Opto-Electronics, Shanxi University,\\
Taiyuan 030006, P.R.China\\}

\begin{abstract}
We propose a new protocol of implementing continuous-variable
quantum state transfer using partially disembodied transport. This
protocol may improve the fidelity at the expense of the introduction
of a semi-quantum channel between the parties, in comparison with
quantum teleportation using the same strength of entanglement.
Depending on the amount of information destroyed in the measurement,
this protocol may be regarded as a teleportation protocol (complete
destruction of input state), or as a $1\rightarrow M$ cloning
protocol (partial destruction), or as a direct transmission (no
destruction). This scheme can be straightforwardly implemented with
the experimentally accessible setup at present.
\end{abstract}


\maketitle

Quantum teleportation is an important protocol in the quantum
information and quantum communication fields, which embodies a basic
law of quantum mechanics - quantum no cloning theorem\cite{one}.
This protocol enables reliable transfer of an arbitrary, unknown
quantum state from one location to another\cite{two}. This transfer
is achieved by utilizing shared quantum entanglement and classical
communication between two locations. The fact that no information
whatsoever is gained on either particle is the reason why quantum
teleportation escapes the verdict of the no-cloning theorem. In
recent years, quantum teleportation has played a central role in
quantum information science and has become an essential tool in
diverse quantum algorithms and protocols\cite{three,four}.

Quantum teleportation was originally developed in the context of the
discrete quantum variables, with the central notion of a single
quantum bit (qubit) as a basic unit. In recent years, theses
concepts have been extended to the domain of continuous variable
(CV), which have attracted a lot of interest and appear to yield
very promising perspectives concerning both experimental
realizations and general theoretical insights, due to relative
simplicity and high efficiency in the generation, manipulation, and
detection of CV state. The first results in this direction concerned
quantum teleportation\cite{five,six} and
experimentally implemented to teleport coherent states with a fidelity $%
{\cal F}=0.58\pm 0.02$\cite{seven}. The fidelity, which quantifies
the success of a teleportation experiment, is defined as ${\cal
F}\equiv \left\langle \psi ^{in}\left| \hat{\varrho}^{out}\right|
\psi ^{in}\right\rangle $, where ``in'' and ``out'' denote the input
and the output state. Later, continuous-variable quantum
teleportation was successfully performed by other
groups\cite{eight,nine1,nine}. Quantum teleportation succeeds when
the fidelity exceeds the classical limit (${\cal F}_c=1/2$ for a
coherent state input) which is the best achievable value without the
use of entanglement\cite{benchmark}. The value of $2/3$ is referred
to as the no-cloning limit, because surpassing this limit warrants
that the teleported state is the best remaining copy of the input
state\cite{ten}. Exceeding this bound would require an EPR
(Einstein-Podolsky-Rosen) channel with more than $3$ $dB$ squeezing.

In the protocol of quantum teleportation\cite{two}, Alice completely
destroys the unknown quantum state by measurement, so divides its
full information into two parts, one purely classical and the other
purely non-classical, and sends them to Bob through two different
channel. In this Letter, we propose a new scheme of quantum state
transfer using partially disembodied transport. Alice divides the
unknown quantum state into two parts, one part is not destroyed and
the other part is destroyed by the joint Bell-state measurement with
a half of the entangled EPR beam. The undestroyed part is displaced
by the measured outcomes and then is transmitted to Bob. This
channel is regarded as the semi-quantum channel, whose features
depends on the amount of information destroyed in the measurement.
Bob will retrieve the initial quantum state under the assistance of
his other half of the entangled EPR beam. Also, no information is
gained in this process, so it is still to obey the no-cloning
theorem. In this case, the fidelity boundary between classical and
quantum transfer depends on the amount of the destroyed information.
The fidelity in this protocol may be improved at the expense of the
introduction of a semi-quantum channel between the parties, in
comparison with quantum teleportation under the same strength of EPR
entanglement. This novel quantum state transfer can be
straightforwardly implemented with the present teleportation
setup\cite{seven,eight,nine1,nine}. Further, we will show the
partially disembodied quantum state transfer relates to the optimal
Gaussian cloning
machine\cite{eleven,twelve,thirteen,forteen,Ferraro,fifteen}.

\begin{figure}[t]
\includegraphics[width=8cm]{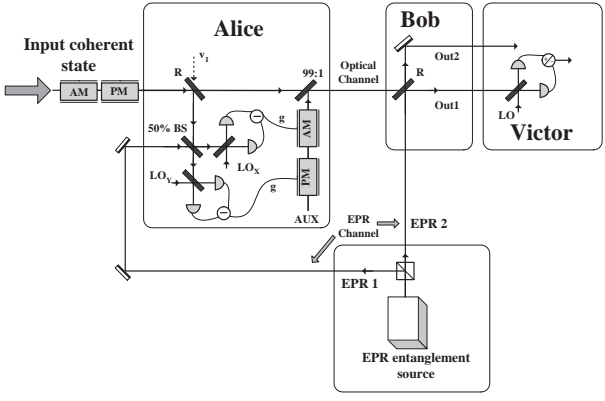}
\caption{Schematic of the CV quantum state transfer using partially
disembodied transport. BS: Beam splitter, g: Electronic gain, LO:
Local oscillator, AM: Amplitude modulator, PM: Phase modulator and
AUX: Auxiliary beam.} \label{fig:als}
\end{figure}

A schematic setup for CV quantum state transfer using partially
disembodied transport is depicted in Fig.1. The quantum states we
consider in this Letter can be described using the electromagnetic
field annihilation operator $\hat{a}=(\hat{X}+i\hat{Y})/2$, which is
expressed in terms of the amplitude $\hat{X}$ and phase $\hat{Y}$
quadrature with the canonical commutation relation $[ \hat{X},\hat{Y}%
] =2i$. Without a loss of generality, the quadrature operators can
be
expressed in terms of a steady state and fluctuating component as $\hat{A}%
=\langle \hat{A}\rangle +\delta \hat{A}$, which have
variances of $V_A=\langle \delta \hat{A}^2\rangle $($\hat{A}%
=\hat{X}$ or $\hat{Y})$. First Alice and Bob share the entangled EPR
beams, one of which is send to a sender Alice and the other is to a
receiver Bob. At Alice's sending station the coherent states of
optical fields are used as input states. The quantum state
$\hat{a}_{in}$ is divided by a beamsplitter with a variable
reflectivity $R$, $0< {R}< 1$. Alice combines the reflected output
state and her entangled beam at a 50/50 beamsplitter, and then
measures $\hat{X}$ and $\hat{Y}$ quadratures by two homodyne
detectors, respectively. Thus Alice only destroys the partial
information of the unknown quantum state from the reflected field by
measurement. Hence, the reflectivity $R$ represents the amount of
the destroyed information of the unknown quantum state. Alice uses
the photocurrents measured by two homodyne detectors to modulate the
amplitude and phase of an auxiliary beam (AUX) via two independent
modulators with a scaling factor $g$\cite{sixteen}. This beam is
then combined at a 99:1 beam splitter with the other undestroyed
part of the the unknown input quantum state, hereby displacing this
part according to measurement outcomes\cite{seven}. In the
Heisenberg representation, the displaced field is expressed by
\begin{eqnarray}
\hat{a}_{disp}&=&(\sqrt{1-R}+\frac g{\sqrt{2}}\sqrt{R})\hat{a}_{in}+(\sqrt{R}-%
\frac g{\sqrt{2}}\sqrt{1-R})\hat{v}_1 \nonumber \\ & &-\frac
g{\sqrt{2}}\hat{b}_{EPR1}^{\dagger } \label{disp}
\end{eqnarray}
where $\hat{v}_1$ refer to the annihilation operator of the vacuum
noise entering the beamsplitter 1, $\hat{b}_{EPR1}$ is the
annihilation operator of EPR beam 1, and $\hat{a}_{disp}$ is the
annihilation operator for the displaced state. The displaced field,
whose vacuum noise $\hat{v}_1$ is cancelled when $g$ is taken to be
$\sqrt{2R/(1-R)}$, is given by
\begin{equation}
\hat{a}_{disp}^c=\frac 1{\sqrt{1-R}}\hat{a}_{in}-\sqrt{\frac
R{1-R}}\hat{b}_{EPR1}^{\dagger }. \label{chann}
\end{equation}
Then the displaced field $\hat{a}_{disp}^c$ is transmitted to a
remote station Bob who reconstructs the unknown quantum state. In
this process the displaced field plays a role of an optical channel
(semi-quantum channel). In the ideal case of perfect EPR
entanglement, one cannot get any information of the input state
$\hat{a}_{in}$
from the optical channel because the amount of quantum noise of $\hat{b}%
_{EPR1}$ is big enough to hide all information of the transmitted
state. After receiving the information from Alice, Bob reconstructs
the unknown state by interfering the transmitted optical field with
his entangled beam at an other beamsplitter with the same
reflectivity $R$. In the absent of losses, the output states from
two output parts of beamsplitter are written by
\begin{eqnarray}
\hat{a}_{out1} &=&\hat{a}_{in}+\sqrt{R}(\hat{b}_{EPR2}-\hat{b}%
_{EPR1}^{\dagger })  \label{out} \\
\hat{a}_{out2} &=&\sqrt{\frac R{1-R}}\hat{a}_{in}-\frac R{\sqrt{1-R
}}\hat{b}_{EPR1}^{\dagger } \nonumber \\ & & -\sqrt{1-R
}\hat{b}_{EPR2} \nonumber
\end{eqnarray}
where $\hat{b}_{EPR2}$ is the annihilation operator of Bob's EPR beam 2. The
EPR entangled beams have the very strong correlation property, such as both
their difference-amplitude quadrature variance $\left\langle \delta (\hat{X}%
_{b_{EPR1}}-\hat{X}_{b_{EPR2}})^2\right\rangle =2e^{-2r}$, and their
sum-phase quadrature variance $\left\langle \delta (\hat{Y}_{b_{EPR1}}+%
\hat{Y}_{b_{EPR2}})^2\right\rangle =2e^{-2r}$, are less than the
quantum noise limit, where $r$ is the squeezing factor. Thus,
normalizing the variance of the vacuum state to unity, the variances
of the output 1 for the amplitude and phase quadratures are
$\left\langle \delta \hat{X}_{out1}^2\right\rangle =\left\langle
\delta \hat{X}_{in}^2\right\rangle +2Re^{-2r}$ and $\left\langle
\delta \hat{Y}_{out1}^2\right\rangle =\left\langle \delta \hat{Y}%
_{in}^2\right\rangle +2Re^{-2r}$. In the case of unity gain, the
fidelity for the Gaussian states is simply given by
\begin{equation}
{\cal F}=\frac 2{\sqrt{\left( 1+\left\langle \delta \hat{X}%
_{out}^2\right\rangle \right) \left( 1+\left\langle \delta \hat{Y}%
_{out}^2\right\rangle \right) }}.
\end{equation}
For the classical case of $r=0$, i.e., the EPR beams were replaced
by uncorrelated vacuum inputs, the fidelity of output 1 is found to
be ${\cal F}_{boun}=1/(R+1)$ which correspond to the fidelity
boundary between classical and quantum transfer as shown in Fig.2.
When Alice and Bob share a EPR entanglement $r>0$, the fidelity of
the output 1 is ${\cal F}_1=1/(1+Re^{-2r})$. It clearly shows that
the fidelity boundary degrades as the amount of the destroyed
information of the unknown quantum state increases, and the quantum
fidelity may be improved at the expense of the introduction of a
semi-quantum channel between the parties when the destroyed
information decreases for a given EPR entanglement. Here, we don't
consider the optimization of the entangled resource. Comparing with
our protocol, the fidelity of continuous variable quantum
teleportation for a given entanglement resource is optimized by
means of the local unitary operations applicable to the entangled
resource itself \cite{opt}. Thus, these methods to optimize the
fidelity of quantum teleportation may be applied to our protocol
directly and improve fidelity further.

\begin{figure}[t]
\includegraphics[width=8cm]{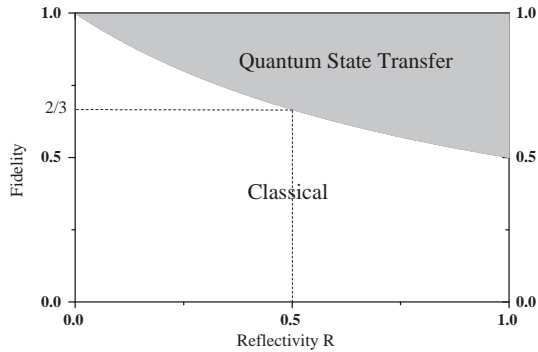}
\caption{The fidelity of quantum state transfer with partially
disembodied transport as a function of the reflectivity $R$ (the
amount of the destroyed information of the unknown quantum state).}
\label{fig2}
\end{figure}

The proposed system has different characteristic and use when $R$
has different value. Let us first consider the case where
$R\rightarrow1$. This means that the unknown quantum state is
destroyed completely. In this case, our scheme is equivalent to the
protocol of all-optical teleportation presented in \cite{seventeen},
and also corresponds to standard quantum teleportation\cite{seven},
in which the displacement operation is implemented by Bob instead of
by Alice. When $R=0.5$, only half information of the unknown quantum
state is destroyed. For the classical case of $r=0$, the fidelity of
both $\hat{a}_{out1}$ and $\hat{a}_{out2}$ is found to be $2/3$.
This scheme corresponds to the optimal fidelity for a $1\rightarrow
2$ symmetric Gaussian cloning machine\cite{fifteen}, which is
experimentally realized recently\cite{eighteen}. When Alice and Bob
share a EPR
entanglement $r>0$, the fidelity of the output 1 is ${\cal F}_1=2/(%
2+e^{-2r})$, which is always larger than $2/3$, and the output 2 is
${\cal F}_2=2/(2+e^{2r})$, which is always smaller than $2/3$. The
result corresponds to the optimal fidelity for a $1\rightarrow 2$
asymmetric
Gaussian cloning machine\cite{forteen,Ferraro}. Thus Bob can achieve ${\cal F}%
_1>2/3$ for an unknown coherent state only by way of shared quantum
entanglement and without requirement of 3 dB squeezing ($e^{-2r}=0.5$%
), but at the expense of the introduction of a semi-quantum channel
between the parties. When $R\rightarrow0$, this means that the
unknown quantum state is transmitted directly to Bob.

Now we show that our proposal may be viewed as a $1\rightarrow M$
optimal Gaussian cloning machine when the reflectivity $R=(M-1)/M$
as shown in Fig.3. At Bob's station, the output 2 is sent together
with $M-2$ ancilla modes through an $(M-2)$-splitter. The ancilla
modes $\hat{v}_{b1},\hat{v}_{b2},.\ldots \hat{v}_{b(M-2)}$ are
vacuum modes. The output states are expressed by
\begin{eqnarray}
\hat{a}_{out1}^{1\rightarrow M} &=&\hat{a}_{in}+\frac{\sqrt{M-1}}{\sqrt{M}}(%
\hat{b}_{EPR2}-\hat{b}_{EPR1}^{\dagger }) \\
\hat{a}_{out2'}^{1\rightarrow M} &=&\hat{a}_{in}-\frac{M-1}{\sqrt{M(M-1)}}%
\hat{b}_{EPR1}^{\dagger } \nonumber \\
&& -\frac 1{\sqrt{M(M-1)}}\hat{b}_{EPR2}+\frac{\sqrt{%
M-2}}{\sqrt{M-1}}\hat{v}_{b1}  \nonumber \\
\cdots \nonumber.
\end{eqnarray}
For the classical case of $r=0$, the fidelity of any outputs is
found to be ${\cal F}_{boun}^{1\rightarrow M}=M/(2M-1)$, which
corresponds to the
optimal fidelity for a $1\rightarrow M$ symmetric Gaussian cloning machine%
\cite{fifteen}. When Alice and Bob share a EPR entanglement $r>0$,
only the
fidelity of the output 1, which is ${\cal F}_1^{1\rightarrow M}=\frac M{%
M+(M-1)e^{-2r}}$, is greater than ${\cal F}_{boun}^{1\rightarrow
M}$. We see the fidelity boundary between classical and quantum
transfer of arbitrary input coherent states depends on how many
identical copies are produced from original one in an optimal
cloner, which lies in $1/2\leq {\cal F}_{boun}\leq 2/3$.

\begin{figure}[t]
\includegraphics[width=7.8cm]{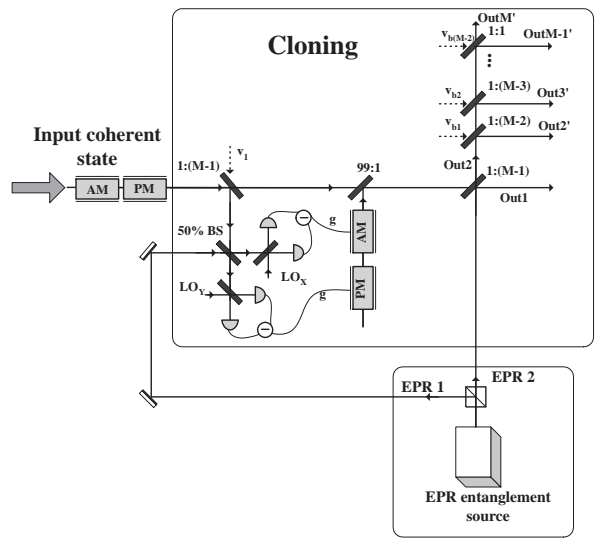}
\caption{Schematic of the quantum state transfer with partially
disembodied transport viewed as $1\rightarrow M$ optimal Gaussian
cloning machine.} \label{fig3}
\end{figure}

Due to the imperfection of EPR Entanglement, some information of the
unknown quantum state may be extracted from optical channel. Here we
will investigate the amount of the information by means of the
signal to noise $(S/N)=V_s/V_n$ where $V_s$ is signal variance and
$V_n$ is the quantum noise variance. For getting the information of
the unknown quantum state from the optical channel, one
simultaneously measures the amplitude and phase quadrature of the
optical channel using a 50\% beamsplitter. Using Eq.(\ref{chann}),
the signal to noise of the measured quadratures are given by
\begin{eqnarray}
(S/N)_{X(Y)}& = &V_{X_{in}(Y_{in})}/[\frac{e^{2r}+e^{-2r}}2\nonumber
\\ &&
+\frac 1{1-R}(1-\frac{%
e^{2r}+e^{-2r}}2)].  \label{sn}
\end{eqnarray}
For the same strength of EPR entanglement Eq.(\ref{sn}) shows that
the information of the unknown quantum state leaked from optical
channel decreases when the amount of the destroyed quantum state
increases.

The another important feature of the optical channel is its
dependence on the propagation losses. If $\eta$ expresses the
transmission losses of the optical channel, the displaced field
transmitted to a remote station Bob is given by
\begin{eqnarray}
\hat{a}_{disp}^{c}& = &\frac 1{\sqrt{1-R}}\hat{a}_{in}+\frac{(\eta
-1)}{\sqrt{R}}\hat{v}_1-\frac{1-\eta(1-R)
}{\sqrt{R(1-R)}}\hat{b}_{EPR1}^{\dagger } \nonumber \\ &&
+\sqrt{1-\eta ^2}\hat{v}_c.
\end{eqnarray}
Here $g=\sqrt{2}(1-\eta(1-R))/\eta\sqrt{R(1-R)}$ ensures that the
displaced field contains the input field $\hat{a}_{in}$ by a factor
of $1/\sqrt{1-R}$. At Bob's station, the output state $1$ is given
by
\begin{eqnarray}
\hat{a}_{out1}& = & \hat{a}_{in}+\sqrt{R}(%
\hat{b}_{EPR2}-\hat{b}_{EPR1}^{\dagger }) \nonumber \\ && -\frac{(1-\eta )(1-R)}{\sqrt{R}}\hat{b%
}_{EPR1}^{\dagger }-\frac{(1-\eta )\sqrt{1-R}}{\sqrt{R}}\hat{v}_1
\nonumber
\\ && +\sqrt{(1-\eta ^2)(1-R)}\hat{v}_c. \label{out1loss}
\end{eqnarray}
Eq.(\ref{out1loss}) clearly shows that the influence of the
propagation losses on the transmitted state decreases as the amount
of the destroyed information of the unknown quantum state increases.
This is due to the fact that the output mode $1$ only contains $1-R$
portion of the displaced field and $R$ portion of entangled beam 2
at the Bob's beamsplitter. Thus the optical channel for
$R\rightarrow 0$ is referred to as the quantum channel, which
transfers the unknown quantum state directly, and the optical
channel for $R\rightarrow 1$ becomes the classical channel, which is
completely independent of the losses. We refer the optical channel
for $0< {R}< 1$ as ``semi-quantum'' channel in order to distinguish
it from the classical and quantum channel.

In conclusion, we have proposed an experimentally feasible scheme of
quantum state transfer using partially disembodied transport for
continuous quantum variables. The fidelity boundary between
classical and quantum transfer of arbitrary input coherent states
depends on the amount of the destroyed information. The fidelity in
this protocol may be improved at the expense of the introduction of
a semi-quantum channel between the parties, in comparison with
quantum teleportation for a given EPR entanglement. We show that the
partially disembodied quantum transfer is related to the
$1\rightarrow M$ optimal Gaussian cloning machine. The optical
channel of partially disembodied transport of an unknown quantum
state is gradually changed from quantum to classical channel with
the increasing reflectivity $R$. This new scheme of implementing
quantum state transfer helps to deepen our understanding of the
properties of quantum communication systems enhanced by EPR
entanglement and its multi-usability and flexibility might have
remarkable application in quantum communication and computation.

$^{\dagger} $Corresponding author's email address:
jzhang74@yahoo.com, jzhang74@sxu.edu.cn

{\bf ACKNOWLEDGMENTS}

J. Zhang thanks T. Zhang, J. Gao, and Q. Pan for the helpful
discussions. This research was supported in part by National Natural
Science Foundation of China (Approval No.60178012, No.60238010), and
Program for New Century Excellent Talents in University.

\end{document}